# Universal Fermi velocity in highly compressed hydride superconductors

Evgeny F. Talantsev[1,2*]

[1]M.N. Miheev Institute of Metal Physics, Ural Branch, Russian Academy of Sciences, 18, S. Kovalevskoy St., Ekaterinburg, 620108, Russia

[2]NANOTECH Centre, Ural Federal University, 19 Mira St., Ekaterinburg, 620002, Russia

*corresponding author's E-mail: evgney.talantsev@imp.uran.ru

**Abstract**

Fermi velocity, $v_F$, is one of the primary characteristics of any conductor, including superconductors. For conductors at ambient pressure several experimental techniques have been developed to measure $v_F$ and, for instance, Zhou *et al* (*Nature* **423** 398 (2003)) reported that high-$T_c$ cuprates exhibit universal nodal Fermi velocity of $v_{F,univ} = (2.7 \pm 0.5) \times 10^5 \frac{m}{s}$. However, there were no experimental techniques applied to measure $v_F$ in highly compressed near-room-temperature superconductors (NRTS), due to experimental challenges. Here to answer a question about the existence of the universal Fermi velocity in NRTS materials, we analyzed full inventory of the ground-state upper critical field data, $B_{c2}(0)$, for these materials and found that this class of superconductors exhibits universal Fermi velocity of $v_{F,univ} = \frac{1}{1.3} \times \left(\frac{2\Delta(0)}{k_B T_c}\right) \times 10^5 \frac{m}{s}$ (where $\Delta(0)$ is ground state amplitude of the energy gap). Due to the ratio of $\frac{2\Delta(0)}{k_B T_c}$ is varying within a narrow arrange of $3.2 \leq \frac{2\Delta(0)}{k_B T_c} \leq 5$, then $v_{F,univ}$ in NRTS materials is in a range of $2.5 \times 10^5 \frac{m}{s} \leq v_{F,univ} \leq 3.8 \times 10^5 \frac{m}{s}$, which is in the same ballpark with its high-$T_c$ cuprates counterpart.

**Universal Fermi velocity in highly compressed hydride superconductors**

## I. Introduction

Since pivotal experimental discovery of first near-room-temperature superconductor (NRTS) $H_3S$ by Drozdov *et al* [1], nearly two dozen of highly compressed hydrogen-rich superconducting phases have been synthesized in binary and ternary systems [2-17]. Experimental studies of NRTS are well supported by first-principles calculations [18-30], however, experimental characterizations of NRTS phases are limited by narrow set of techniques, which can be applied for materials inside of diamond anvil cell (DAC) [25-27]. These techniques are X-ray diffraction (XRD) phase analysis, Raman spectroscopy and magnetoresistance measurements [31-35]. In some advanced experiments, Hall effect measurements can be also performed [31]. Based on this, only two characteristic values of the superconducting state of the NRTS phases are commonly extracted from the experimental data, which are the transition temperature, $T_c$, and the extrapolated value for the ground state upper critical field, $B_{c2}(0)$ or the ground state superconducting coherence length, $\xi(0)$, which can be derived from the Ginzburg-Landau [36] expression:

$$\xi(0) = \sqrt{\frac{\phi_0}{2\pi B_{c2}(0)}} \qquad (1)$$

where $\phi_0 = \frac{h}{2e}$ is the superconducting flux quantum, $h$ is Planck constant and $e$ is electron electric charge.

Other important parameters of the NRTS materials, from which we can mention the Fermi velocity, $v_F$, cannot be measured to date, due to challenging experimental problems associated with measurement of this value for samples inside of DAC. However, considering that all NRTS superconductors are hydrides, there is an expectation, that these materials can exhibit universal Fermi velocity, $v_{F,univ}$, as the one was discovered in cuprates, $v_{F,univ} =$

( $2.7 \pm 0.5$ ) × $10^5 \frac{m}{s}$ (which was reported by Zhou *et al* [37]). In Figure 1 we showed the dataset reported by Zhou *et al* [37].

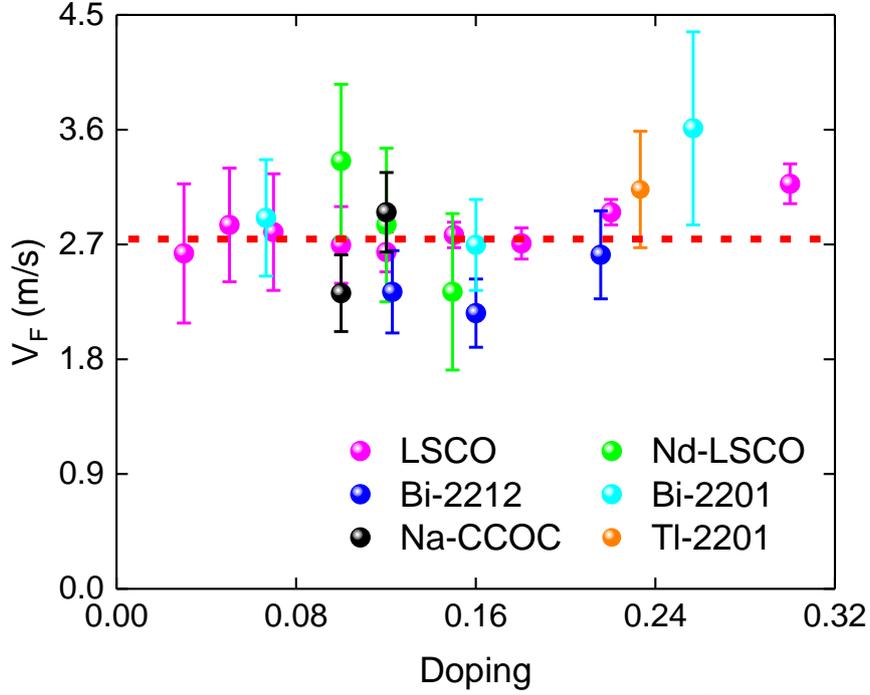

**Figure 1.** Universal nodal Fermi velocity, $v_{F,univ} = ( 2.7 \pm 0.5 ) \times 10^5 \frac{m}{s}$, for cuprate superconductors. Raw data reported by Zhou *et al* [37]. Data presented for (La$_{2-x}$Sr$_x$)CuO$_4$ (LSCO), (La$_{2-x-y}$Nd$_y$Sr$_x$)CuO$_4$ (Nd-LSCO), Bi$_2$Sr$_2$CaCu$_2$O$_8$ (Bi-2212), Bi$_2$Sr$_2$CuO$_6$ (Bi-2201), (Ca$_{2-x}$Na$_x$)CuO$_2$Cl$_2$ (Na-CCOC), and Tl$_2$Ba$_2$CuO$_6$ (Tl-2201).

Partial theoretical background for the quest for universal Fermi velocity in NRTS is based from one hand on recent understanding [38] that sulphur in H$_3$S is an analogue to the oxygen in cuprates, and from other hand that highly compressed hydrides are nicely added in main global scaling laws for superconductors [39-42].

Here, we reported the result of our search for universal Fermi velocity in NRTS materials which was based on the analysis of full inventory of the ground state upper critical field, $B_{c2}(0)$ in these materials. In the result, we found that universal Fermi velocity, $v_{F,univ}$, does exist in NRTS materials, and the one obeys the empirical law:

$$v_{F,univ} = \frac{1}{1.3} \times \frac{2\Delta(0)}{k_B T_c} \times 10^5 \left( \frac{m}{s} \right) \qquad (2)$$

where $k_B$ is the Boltzmann constant, and $\Delta(0)$ is the ground state superconducting energy gap.

## II. Approach description

In Bardeen-Cooper-Schieffer (BCS) theory of superconductivity [43] the ground state coherence length, $\xi(0)$, and the amplitude of the ground state energy gap, $\Delta(0)$, are linked through the expression:

$$\xi(0) = \frac{\hbar v_F}{\pi \Delta(0)} \qquad (3)$$

where $\hbar$ is reduced Planck constant. BCS theory also has a dimensionless ratio:

$$\alpha = \frac{2\Delta(0)}{k_B T_c} \qquad (4)$$

By substituting Eqs. 3,4 in Eq. 1, one can get a dependence of the ground state upper critical field vs the transition temperature:

$$B_{c2}(0) = \left[\frac{\pi \phi_0 k_B^2}{8\hbar^2}\right] \times \frac{\alpha^2}{v_F^2} \times T_c^2 \qquad (5)$$

where the multiplicative pre-factor in square brackets is a constant:

$$A = \left[\frac{\pi \phi_0 k_B^2}{8\hbar^2}\right] = 1.38 \times 10^7 \frac{T \times m^2}{s^2 \times K^2} \qquad (6)$$

Thus, if hydrogen-rich superconductors exhibit universal Fermi velocity, $v_{F,univ}$, the fit of full inventory of $B_{c2}(0)$ vs $T_c$ dataset to the equation of:

$$B_{c2}(0) = A \times f \times T_c^{\beta} \qquad (7)$$

where $\beta$ and $f = \frac{\alpha^2}{v_F^2}$ are free fitting parameters, should reveal that:

$$\beta \cong 2 \qquad (8)$$

and if this is a case, then universal Fermi velocity, $v_{F,univ}$, can be calculated from deduced free-fitting parameter $f$:

$$v_{F,univ} = \frac{\alpha}{\sqrt{f}} = \frac{1}{\sqrt{f}} \times \frac{2\Delta(0)}{k_B T_c} \qquad (9)$$

It should be noted, that $\alpha = \frac{2\Delta(0)}{k_B T_c}$ in highly-compressed hydrogen-rich superconductors is varying within a range [8,12,27,42,44-48]:

$$3.2 \leq \frac{2\Delta(0)}{k_B T_c} \leq 5 \qquad (10)$$

(where the lower limit is the value deduced from experiment [42,44,48], while the upper limit is based on many results reported by the first-principles calculations, which always predict $4.3 \leq \frac{2\Delta(0)}{k_B T_c}$ [8,12,27,45-47] in NRTS materials).

### III. Extrapolation model for the ground state upper critical field

Eq. 7 has the ground state upper critical field, $B_{c2}(0)$, as dependent variable. However, it is important to note that this value can be determined by the use of extrapolative models [49-53] which use experimental $B_{c2}(T)$ data measured at high reduced temperatures, $\frac{T}{T_c}$. Primary reason, why there is a necessity for extrapolative models, is that all highly-compressed hydrogen-rich superconductors have $B_{c2}(T \to 0\,K) > 20\,T$, which cannot be measured by conventional PPMS systems (manufactured by Quantum Design) where the highest magnetic field is limited by $B_{appl} = 9\text{-}16$ Tesla (depends on the model). It should be also stressed, that $B_{c2}(T \to 0\,K)$ for NRTS compounds of $H_3S$, $LaH_{10}$, $YH_6/YH_9$ and $(La,Y)H_{10}$ are so high, that even experimental data measured at world-top quasi-DC magnetic field facility [31,54] only covers the range of reduced temperatures $\frac{1}{2} \leq \frac{T}{T_c}$

From several available extrapolative $B_{c2}(T)$ models [49-53] in this paper we used analytical approximative expression for Werthamer-Helfand-Hohenberg (WHH) theory [55,56], which was proposed by Baumgartner $et\ al$ [53] (and, thus, Eq. 11 we will designate as B-WHH model):

$$B_{c2}(T) = \frac{1}{0.693} \times \frac{\phi_0}{2\pi\xi^2(0)} \times \left( \left(1 - \frac{T}{T_c}\right) - 0.153 \times \left(1 - \frac{T}{T_c}\right)^2 - 0.152 \times \left(1 - \frac{T}{T_c}\right)^4 \right) \qquad (11)$$

where $\xi(0)$ and $T_c \equiv T_c(B=0)$ are two free fitting parameters. Eq. 11 [53] was initially proposed to extrapolate $B_{c2}(T)$ data for neutron-irradiated $Nb_3Sn$ alloys, and recently several research groups found that Eq. 11 is a good approximated tool for a variety of superconducting materials [4,57-62]. Based on this, in current study we used Eq. 11 as a good, robust and simple analytical tool to extrapolate $B_{c2}(T)$ curve on low temperature/high field region [4,57-62], because, as we mentioned above, $B_{c2}(T)$ datasets for NRTS superconductors are measured only at high reduced temperatures, $\frac{1}{2} \leq \frac{T}{T_c}$, because of experimental limitations.

There is a need to describe the criterion to extracting $B_{c2}(T)$ datasets from experimentally measured $R(T,B_{appl})$ curves. There are several criteria for the $T_c$, $B_{c2}(T)$ and $T_c(B_{appl})$ definition, which for the case of NRTS discussed recently in Ref. 63. In the result we found [63,64] that the best match between the electron-phonon coupling constant $\lambda_{e-ph}$ extracted from $R(T,B_{appl}=0)$ curves and $\lambda_{e-ph}$ computed by first principles calculation is when $T_c$ is defining at as low as practically possible fraction of $R(T)/R_{norm}$ (where $R_{norm}$ is the normal state resistance just above the transition). By analysing full inventory of $R(T,B_{appl})$ data for NRTS materials herein, we came to conclusion that due to noise/slope issues of real-world $R(T,B_{appl})$ curves and a fact that highly-compressed superhydrides contained several superconducting phases the appropriate criterion, which we used in this study is:

$$\frac{R(T,B_{appl})}{R(T_c^{onset},B_{appl})} = 0.05 \tag{12}$$

## IV. Results

### 4.1. Unannealed highly-compressed sulphur hydride

In the first paper on NRTS superconductors, Drozdov *et al* [1] reported $R(T,B_{appl})$ data for unannealed highly-compressed sulphur hydride ($P = 155$ GPa) in their Figure 3(a). By using the criterion of Eq. 12 (which is $R(T, B_{appl})_{criterion} = 23\ m\Omega$ for given $R(T, B_{appl})$ curves showed in bottom insert in Figure 3(a) in Ref. 1), we extracted $B_{c2}(T)$ dataset for this sample, which is shown in Fig. 2. Because this $B_{c2}(T)$ dataset covers significant part of full temperature range, $0\ K < T \leq T_c$, there was no need to use extrapolative fit and instead we fitted this dataset to the model [48], which allows to deduce $\Delta(0)$, $\frac{2\Delta(0)}{k_B T_C}$, $\Delta C/C$ (which is the relative jump in electronic specific heat at $T_c$):

$$B_{c2}(T) = \frac{\phi_0}{2\pi\xi^2(0)} \times \left(\frac{1.77 - 0.43\left(\frac{T}{T_C}\right)^2 + 0.07\left(\frac{T}{T_C}\right)^4}{1.77}\right)^2 \times \left[1 - \frac{1}{2k_BT} \times \int_0^\infty \frac{d\varepsilon}{cosh^2\left(\frac{\sqrt{\varepsilon^2 + \Delta^2(T)}}{2k_BT}\right)}\right] \qquad (13)$$

where temperature dependent superconducting gap, $\Delta(T)$, is given by [65,66]:

$$\Delta(T) = \Delta(0) \times \tanh\left[\frac{\pi k_B T_c}{\Delta(0)} \times \sqrt{\eta \times \frac{\Delta C}{C} \times \left(\frac{T_c}{T} - 1\right)}\right] \qquad (14)$$

where $\eta$ = 2/3 for $s$-wave superconductors.

Eqs. 13,14 were used to extract $\xi(0)$, $\Delta(0)$, $T_c$ and $\frac{\Delta C}{C}$ from $B_{c2}(T)$ datasets in a variety of superconductors, for instance, for two highly-compressed hydrides phases of $H_3S$ [48] and of $SnH_{12}$ [42], $V_3Si$ [67], and several iron-based superconductors [67]. However, it should be stressed that the approach (i.e. Eq. 13,14) is only applicable for $B_{c2}(T)$ datasets defined by Eq. 12 or by stricter criterion.

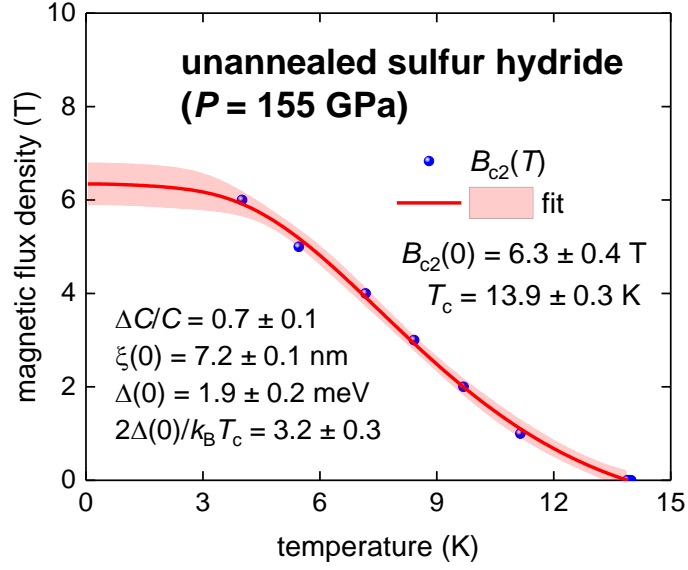

**Figure 2.** The upper critical field data, $B_{c2}(T)$, and data fit to Eqs. 13,14 for unannealed highly-compressed sulphur hydride ($P$ = 190 GPa). Raw $R(T,B_{appl})$ dataset reported by Drozdov *et al* [1]. Deduced values are shown in the figure. 95% confidence bands are shown by a pink shaded area. Fit quality is $R = 0.9985$.

One of the most important deduced parameters, $\alpha = \frac{2\Delta(0)}{k_B T_c} = 3.2 \pm 0.3$, is in remarkable agreement with counterpart values deduced for highly-compressed annealed $H_3S$ ($P$ = 155-160 GPa), $\frac{2\Delta(0)}{k_B T_c} = 3.20 \pm 0.02$ [44] and $\frac{2\Delta(0)}{k_B T_c} = 3.55 \pm 0.31$ [48], and for highly-compressed annealed $SnH_{12}$ ($P$ = 190 GPa), $\frac{2\Delta(0)}{k_B T_c} = 3.28 \pm 0.18$ [42]. Deduced $\frac{\Delta C}{C} = 0.7 \pm 0.1$ is also below the weak-coupling limit of BCS theory $\frac{\Delta C}{C} = 1.43$, as its counterpart in the annealed $H_3S$ material, $\frac{\Delta C}{C} = 1.2 \pm 0.3$ [48]. It should be mentioned that to deduce $\frac{\Delta C}{C}$ with higher accuracy requires more $B_{c2}(T)$ datapoints, especially at $T \sim T_c$. Deduced $B_{c2}(0)$ and $T_c$ are included in Table I.

### 4.2. Annealed highly-compressed hydrides

Reported $R(T,B_{appl})$ datasets for several annealed highly-compressed hydrides were processed by utilizing Eq. 12 to extract $B_{c2}(T)$ datasets. Obtained datasets were fitted to Eq. 11 and deduced values included in Table I. These materials are:

1.  Sulphur superhydride, $H_3S$ ($P$ = 155, 160 GPa), for which raw data reported by Mozaffari *et al* [31]. Fits are shown in Figure S1.

2.  Cerium superhydride, $CeH_n$ ($P$ = 88, 137, 139 GPa), for which raw data reported by Chen *et al* [12]. Fits are shown in Figure S2.

3.  Lanthanum superhydride, $LaH_{10}$ ($P$ = 120, 136 GPa), for which raw data reported by Sun *et al* [54]. Fits are shown in Figure S3.

4.  Yttrium superhydride/superdeiteride, $YH_6/YD_6$ ($P$ = 172, 200 GPa), for which raw data reported by Troyan *et al* [4]. Fits are shown in Figure S4.

5.  Lanthanum-yttrium superhydride, $(La,Y)H_{10}$ ($P$ = 182, 183, 186 GPa), for which raw data reported by Semenok *et al* [8]. Fits are shown in Figure S5.

6.  Tin superhydride, $SnH_{12}$ ($P$ = 190 GPa), for which raw data reported by Hong *et al* [11]. Fits are shown in Figure S6.

7.  Thorium superhydrides, $ThH_9$ and $ThH_{10}$ ($P$ = 170 GPa), for which raw data reported by Semenok *et al* [16]. Fits are shown in Figure S7.

### 4.3. Analysis of $B_{c2}(0)$ vs $T_c$ for superhydride phases

All deduced $B_{c2}(0)$ and $T_c$ values for superhydride phases are collected in Table I, where we also added data for $Th_4H_{15}$ phase reported by Satterthwaite and Toepke [68].

**Table I.** Deduced $B_{c2}(0)$ and $T_c$ values for hydrogen-rich superconductors for which raw $R(T,B_{appl})$ data are available to date (references are included).

| Phase and Data Source | Figure No. | Pressure (GPa) | $T_c$ (K) | $\Delta T_c$ (K) | $B_{c2}(0)$ (T) | $\Delta B_{c2}(0)$ (T) |
|---|---|---|---|---|---|---|
| Unannealed sulphur hydride (Fig. 3(a) in Ref. 1) | 1 | 155 | 13.9 | 0.3 | 6.3 | 0.4 |
| Annealed $H_3S$ (Fig. 3 in Ref. 31) | 2(a) | 155 | 185 | 2 | 98.8 | 1.2 |
| Annealed $H_3S$ (Figs. S1,S2 in Ref. 31) | 2(b) | 155 | 196.1 | 0.6 | 71.1 | 1.1 |
| Annealed $H_3S$ (Fig. 3 in Ref. 31) | 2(c) | 160 | 143.9 | 1.4 | 59.2 | 2.3 |
| Annealed $CeH_9$ (Fig. S7(a) in Ref. 12) cooling | 3(a) | 88 | 38.8 | 0.4 | 16.5 | 1 |
| Annealed $CeH_9$ (Fig. 1(c) in Ref. 12) warming | 3(b) | 139 | 88.6 | 0.3 | 22.2 | 0.7 |
| Annealed $CeH_9$ (Fig. 1(d) in Ref. 12) cooling | 3(c) | 137 | 81.9 | 0.7 | 18.4 | 0.7 |
| Annealed $CeH_9$ (Fig. 1(d) in Ref. 12) warming | 3(d) | 137 | 82.7 | 0.7 | 18.7 | 0.6 |
| Annealed $LaH_{10}$ (Fig. 3(a) in Ref. 47) | 4(a) | 120 | 174.8 | 0.8 | 90 | 3 |
| Annealed LaH$_{10}$ (Fig. 3(b) in Ref. 47) | 4(b) | 136 | 206.2 | 0.8 | 136 | 3 |
| Annealed YD$_6$ (Fig. S13(a) in Ref. 4) | 5(a) | 172 | 157.7 | 0.2 | 124.9 | 2.4 |
| Annealed YH$_6$ (Fig. S16(c) in Ref. 4) | 5(b) | 200 | 206.2 | 0.2 | 97.2 | 1.4 |
| Annealed (La,Y)H$_{10}$ (Fig. S27(b) in Ref. 8) | 6(a) | 183 | 203.5 | 0.2 | 101.6 | 1.8 |
| Annealed (La,Y)H$_{10}$ (Fig. S28(a) in Ref. 8) | 6(b) | 182 | 234 | 0.1 | 135.8 | 1.5 |
| Annealed (La,Y)H$_{10}$ (Fig. S28(a) in Ref. 8) | 6(c) | 186 | 234.5 | 0.1 | 134 | 1 |
| Annealed SnH$_{12}$ (Fig. 4(a) in Ref. 11) cooling | 7(a) | 190 | 62.8 | 0.4 | 9 | 0.2 |
| Annealed SnH$_{12}$ (Fig. 4(a) in Ref. 11) warming | 7(b) | 190 | 64.1 | 0.5 | 8.9 | 0.2 |
| Annealed ThH$_9$ (Fig. 4(a) in Ref. 16) | 8(a) | 170 | 151.2 | 1.5 | 32 | 0.9 |
| Annealed ThH$_{10}$ (Fig. 4(a) in Ref. 16) | 8(b) | 170 | 150.6 | 0.4 | 43.4 | 0.6 |
| Th$_4$H$_{15}$ (Ref. 68) | | ambient | 8.2 | 0.15 | 2.75 | 0.25 |

Full dataset from Table I is shown in Figure 3 together with the fit to Eq. 7. Despite a fact that this dataset has a large scattering, it can be seen in Figure 3(a), that free-fitting power-law exponent, $\beta = 2.07 \pm 0.14$, is practically undistinguishable from expected $\beta \equiv 2$ value (Eq. 5). For the case when $\beta$ is free-fitting parameter (Figure 9(a)), deduced $f = (1.19 \pm 0.90) \times 10^{-10} \frac{s^2}{m^2}$ has a large uncertainty. However, when $\beta$ is fixed to 2 (Figure 3(b)), free-fitting parameter $f$ can be deduced with high accuracy:

$$f = \frac{\alpha^2}{v_{F,univ}^2} = (1.68 \pm 0.08) \times 10^{-10} \frac{s^2}{m^2} \qquad (15)$$

From Equation 15, one can obtain:

$$v_{F,univ} = \frac{\alpha}{(1.30 \pm 0.03)} \times 10^5 \ \frac{m}{s} \cong \frac{1}{1.3} \times \frac{2\Delta(0)}{k_B T_c} \times 10^5 \ \frac{m}{s} \qquad (16)$$

$$2.5 \times 10^5 \ \frac{m}{s} \lesssim v_{F,univ} \lesssim 3.8 \times 10^5 \ \frac{m}{s} \qquad (17)$$

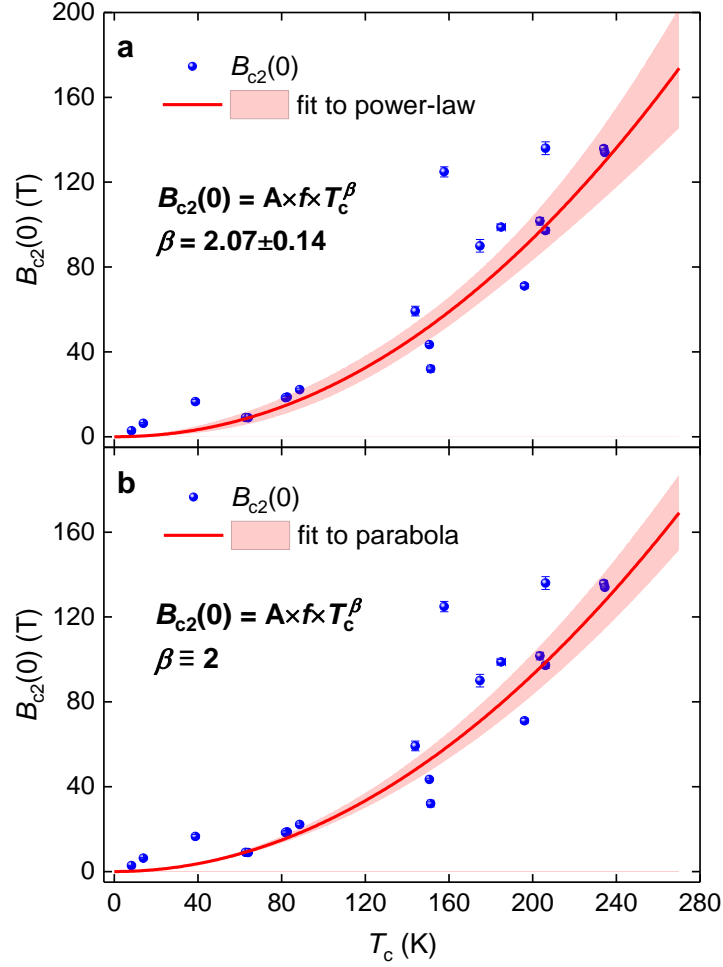

**Figure 3.** Total $B_{c2}(0)$ vs $T_c$ dataset for hydrogen-rich superconductors deduced in this work (Table I) and data fit to (a) Eq. 7 and (b) Eq. 5. (a) – free-fitting $\beta = 2.07 \pm 0.14$ and $f = (1.19 \pm 0.90) \times 10^{-10} \frac{s^2}{m^2}$, fit quality is $R = 0.9361$. (b) – $\beta = 2.0$ (fixed) and free-fitting $(1.68 \pm 0.08) \times 10^{-10} \frac{s^2}{m^2}$, fit quality is $R = 0.9354$.

Deduced $v_{F,univ}$ for hydrogen-rich superconductors (Eq. 16) is at the same ballpark as its counterpart for high-$T_c$ cuprates $v_{F,univ} = (2.7 \pm 0.5) \times 10^5 \frac{m}{s}$ [37], if one takes into account Eq. 10.

## V. Conclusions

In this study we proposed that hydrogen-rich superconductors, including near-room-temperature superconductors, form distinguished subclass of superconducting materials, which exhibits universal Fermi velocity, $v_F$, which is given by empirical expression of:

$v_{F,univ} = \frac{1}{1.3} \times \frac{2\Delta(0)}{k_B T_c} \times 10^5 \frac{m}{s}$. Considering that the gap-to-transition temperature ratio is varying within $\frac{2\Delta(0)}{k_B T_c}$

**Acknowledgement**

The author thanks financial support provided by the Ministry of Science and Higher Education of Russia (theme "Pressure" No. AAAA-A18-118020190104-3) and by Act 211 Government of the Russian Federation, contract No. 02.A03.21.0006.

**Data Availability Statement**

No new data were created or analysed in this study. Data sharing is not applicable to this article.

**References**

[1] A. P. Drozdov, M.I. Eremets, I.A. Troyan, V. Ksenofontov, S. I. Shylin, Conventional superconductivity at 203 kelvin at high pressures in the sulfur hydride system *Nature* **525**, 73-76 (2015)

[2] A. P. Drozdov, *et al* Superconductivity at 250 K in lanthanum hydride under high pressures *Nature* **569**, 528-531 (2019)

[3] M. Somayazulu, *et al* Evidence for superconductivity above 260 K in lanthanum superhydride at megabar pressures *Phys. Rev. Lett.* **122** 027001 (2019)

[4] I. A. Troyan, *et al* Anomalous high-temperature superconductivity in $YH_6$ *Adv. Mater.* **33** 2006832 (2021)

[5] P. P. Kong, *et al* Superconductivity up to 243 K in yttrium hydrides under high pressure *Nature Communications* **12**, 5075 (2021)

[6] L. Ma, *et al* Experimental observation of superconductivity at 215 K in calcium superhydride under high pressure (arXiv:2103.16282)

[7] Z. W. Li, *et al* Superconductivity above 200 K observed in superhydrides of calcium (arXiv:2103.16917)

[8] D. V. Semenok, *et al* Superconductivity at 253 K in lanthanum–yttrium ternary hydrides *Materials Today* **48**, 18-28 (2021)

[9] A. P. Drozdov, M. I. Eremets and I. A. Troyan, Superconductivity above 100 K in $PH_3$ at high pressures (arXiv:1508.06224) [26] D. Zhou, *et al*. Superconducting praseodymium superhydrides *Sci. Adv.* **6**, eaax6849 (2020)

[10] T. Matsuoka, *et al*. Superconductivity of platinum hydride *Phys. Rev. B* **99**, 144511 (2019)

[11] F. Hong, *et al*. Superconductivity at ~70 K in tin hydride $SnH_x$ under high pressure *Materials Today Physics* **22** 100596 (2022)

[12] W. Chen, D. V. Semenok, X. Huang, H. Shu, X. Li, D. Duan, T. Cui and A. R. Oganov, High-temperature superconducting phases in cerium superhydride with a $T_c$ up to 115 K below a pressure of 1 Megabar, *Phys. Rev. Lett.* **127**, 117001 (2021)

[13] M. Sakata, *et al*. Superconductivity of lanthanum hydride synthesized using $AlH_3$ as a hydrogen source *Superconductor Science and Technology* **33** 114004 (2020)

[14] W. Chen, *et al*. High-pressure synthesis of barium superhydrides: Pseudocubic $BaH_{12}$ *Nature Communications* **12**, 273 (2021)

[15] M. A. Kuzovnikov and M. Tkacz, High-pressure synthesis of novel polyhydrides of Zr and Hf with a $Th_4H_{15}$-type structure *J. Phys. Chem. C* **123**, 30059–30066 (2019)

[16] D. V. Semenok, *et al*, Superconductivity at 161 K in thorium hydride $ThH_{10}$: synthesis and properties *Mater. Today* **33**, 36–44 (2020)

[17] N. Wang, *et al* A low-$T_c$ superconducting modification of $Th_4H_{15}$ synthesized under high pressure *Superconductor Science and Technology* **34** 034006 (2021)

[18] H. Xie, *et al*. Superconducting zirconium polyhydrides at moderate pressures *J. Phys. Chem. Lett.* **11**, 646–651 (2020)

[19] S. Mengyao, *et al*. Superconducting $ScH_3$ and $LuH_3$ at megabar pressures *Inorganic Chemistry* (2021); in press: https://doi.org/10.1021/acs.inorgchem.1c01960

[20] J. Chen, *et al*. Computational design of novel hydrogen-rich YS–H compounds *ACS Omega* **4** 14317-14323 (2019)

[21] J. A. Alarco, P. C. Talbot and I. D. R. Mackinnon, Identification of superconductivity mechanisms and prediction of new materials using Density Functional Theory (DFT) calculations *J. Phys.: Conf. Ser.* **1143**, 012028 (2018)

[22] D. V. Semenok, A. G. Kvashnin, I. A. Kruglov, and A. R. Oganov, Actinium hydrides $AcH_{10}$, $AcH_{12}$, and $AcH_{16}$ as high-temperature conventional superconductors *J. Phys. Chem. Lett.* **9** 1920-1926 (2018)

[23] C. J. Pickard, I. Errea, and M. I. Eremets, Superconducting hydrides under pressure *Annual Review of Condensed Matter Physics* **11**, 57-76 (2020).

[24] J. A. Flores-Livas, L. Boeri, A. Sanna, G. Profeta, R. Arita, M. Eremets. A perspective on conventional high-temperature superconductors at high pressure: Methods and materials. *Physics Reports* **856**, 1-78 (2020).

[25] A. Goncharov, Phase diagram of hydrogen at extreme pressures and temperatures; updated through 2019 (Review article), *Low Temperature Physics* **46**, 97 (2020).

[26] E. Gregoryanz, C. Ji, P. Dalladay-Simpson, B. Li, R. T. Howie, and H.-K. Mao, Everything you always wanted to know about metallic hydrogen but were afraid to ask. *Matter and Radiation at Extremes* **5**, 038101 (2020).

[27] L. Boeri, *et al*. The 2021 room-temperature superconductivity roadmap. *Journal of Physics: Condensed Matter* (2021), accepted manuscript: https://doi.org/10.1088/1361-648X/ac2864

[28] X. Zhang, Y. Zhao, G. Yang, Superconducting ternary hydrides under high pressure *WIREs Computational Molecular Science* (2021) in press: https://doi.org/10.1002/wcms.1582

[29] M. Dogan and M. L. Cohen, Anomalous behaviour in high-pressure carbonaceous sulfur hydride *Physica C* 1353851 (https://doi.org/10.1016/j.physc.2021.1353851) (2021)

[30] T. Wang, *et al*. Absence of conventional room temperature superconductivity at high pressure in carbon doped $H_3S$ *arXiv*:2104.03710 (2021)

[31] S. Mozaffari, *et al* Superconducting phase diagram of $H_3S$ under high magnetic fields *Nat. Commun.* **10** 2522 (2019)

[32] V. S. Minkov, V. B. Prakapenka, E. Greenberg, M. I. Eremets, Boosted $T_c$ of 166 K in superconducting $D_3S$ synthesized from elemental sulfur and hydrogen *Angew. Chem. Int. Ed*, **59**, 18970-18974 (2020)

[33] R. Matsumoto, *et al.* Electrical transport measurements for superconducting sulfur hydrides using boron-doped diamond electrodes on bevelled diamond anvil *Superconductor Science and Technology* **33** 124005 (2020)

[34] D. Laniel, *et al.* Novel sulfur hydrides synthesized at extreme conditions *Phys. Rev. B* **102**, 134109 (2020)

[35] X. Huang, *et al.* High-temperature superconductivity in sulfur hydride evidenced by alternating-current magnetic susceptibility *National Science Review* **6** 713-718 (2019)

[36] V. L. Ginzburg and L.D. Landau, Zh. Eksp. Teor. Fiz. **20**, 1064 (1950).

[37] X. J. Zhou, *et al*, High-temperature superconductors: Universal nodal Fermi velocity *Nature* **423** 398 (2003).

[38] D. K. Sunko, High-temperature superconductors as ionic metals *Journal of Superconductivity and Novel Magnetism* **33**, 27-33 (2020)

[39] D. R. Harshman and A. T. Fiory, High-$T_c$ superconductivity in hydrogen clathrates mediated by Coulomb interactions between hydrogen and central-atom electrons *Journal of Superconductivity and Novel Magnetism* **33**, 2945-2961 (2020)

[40] D. R. Harshman and A. T. Fiory, The superconducting transition temperatures of C-S-H based on inter-sublattice S−H4-tetrahedron electronic interactions *Journal of Applied Physics* **131**, 015105 (2022)

[41] Y. J. Uemura, Bose-Einstein to BCS crossover picture for high-Tc cuprates *Physica C* **282–287**, 194-197 (1997)

[42] E. F. Talantsev, Comparison of highly-compressed $C2/m$-$SnH_{12}$ superhydride with conventional superconductors *J. Phys.: Condens. Matter* **33** 285601 (2021)

[43] J. Bardeen, L N Cooper, and J R Schrieffer, Theory of superconductivity *Phys. Rev.* **108**, 1175-1204 (1957)

[44] E. F. Talantsev, W. P. Crump, J. G. Storey, J. L. Tallon, London penetration depth and thermal fluctuations in the sulphur hydride 203 K superconductor *Annalen der Physik* **529** 1600390 (2017)

[45] I. Errea, *et al*, Quantum crystal structure in the 250-kelvin superconducting lanthanum hydride *Nature* **578** 66-69 (2020)

[46] C. Heil, S. di Cataldo, G. B. Bachelet and L. Boeri, Superconductivity in sodalite-like yttrium hydride clathrates *Physical Review B* **99** 220502(R) (2019)

[47] J. A. Camargo-Martínez, *et al*, The higher superconducting transition temperature $T_c$ and the functional derivative of $T_c$ with $\alpha^2F(\omega)$ for electron–phonon superconductors *J. Phys.: Condens. Matter* **32** 505901 (2020)

[48] E. F. Talantsev, Classifying superconductivity in compressed $H_3S$ *Modern Physics Letters B* **33**, 1950195 (2019)

[49] C. J. Gorter and H. Casimir, On supraconductivity I *Physica* **1** 306-320 (1934)

[51] C. K. Jones, J. K. Hulm and B. S. Chandrasekhar, Upper critical field of solid solution alloys of the transition elements *Rev. Mod. Phys.* **36** 74- (1964)

[52] L. P. Gor'kov, The critical supercooling field in superconductivity theory, *Sov. Phys. JETP* **10** 593-599 (1960)

[53] T. Baumgartner, M. Eisterer, H. W. Weber, R. Fluekiger, C. Scheuerlein, L. Bottura, Effects of neutron irradiation on pinning force scaling in state-of-the-art $Nb_3Sn$ wires *Supercond. Sci. Technol.* **27** 015005 (2014)

[54] D. Sun, *et al*., High-temperature superconductivity on the verge of a structural instability in lanthanum superhydride *Nature Communications* **12** 6863 (2021)

[55]  E. Helfand and N. R. Werthamer, Temperature and purity dependence of the superconducting critical field, $H_{c2}$. II. *Phys. Rev.* **147** 288-294 (1966)

[56]  N. R. Werthamer, E. Helfand and P. C. Hohenberg, Temperature and purity dependence of the superconducting critical field, $H_{c2}$. III. Electron spin and spin-orbit effects *Phys. Rev.* **147**, 295-302 (1966)

[57]  H. Ninomiya, *et al*, Superconductivity in a scandium borocarbide with a layered crystal structure *Inorg. Chem.* **58** 15629-15636 (2019)

[58]  H. Xie, *et al*, Superconducting zirconium polyhydrides at moderate pressures *The Journal of Physical Chemistry Letters* **11** 646-651 (2020)

[59]  W. Zhang, *et al*, A New superconducting 3R-WS$_2$ phase at high pressure *J. Phys. Chem. Lett.* **12** 3321-3327 (2021)

[60]  M. Scuderi, et al, Nanoscale analysis of superconducting Fe(Se,Te) epitaxial thin films and relationship with pinning properties *Scientific Reports* **11** 20100 (2021)

[61]  KeYuan Ma, *et al*, Group-9 transition-metal suboxides adopting the filled-Ti$_2$Ni structure: A class of superconductors exhibiting exceptionally high upper critical fields *Chem. Mater.* **33** 8722–8732 (2021)

[62]  M. Boubeche, *et al*, Enhanced superconductivity with possible re-appearance of charge density wave states in polycrystalline Cu$_{1-x}$Ag$_x$Ir$_2$Te$_4$ alloys *Journal of Physics and Chemistry of Solids* **163** 110539 (2022)

[63]  E. F. Talantsev, Advanced McMillan's equation and its application for the analysis of highly-compressed superconductors *Superconductor Science and Technology* **33** 094009 (2020)

[64]  E. F. Talantsev, The electron-phonon coupling constant and the Debye temperature in polyhydrides of thorium, hexadeuteride of yttrium, and metallic hydrogen phase III *Journal of Applied Physics* **130** 195901 (2021)

[65]  F. Gross, *et al.* Anomalous temperature dependence of the magnetic field penetration depth in superconducting UBe$_{13}$. *Z. Phys. B* **64** 175-188 (1986)

[66]  F. Gross-Alltag, B. S. Chandrasekhar, D. Einzel, P. J. Hirschfeld and K. Andres, London field penetration in heavy fermion superconductors *Z. Phys. B* **82** 243-255 (1991)

[67]  E. F.Talantsev, In-plane *p*-wave coherence length in iron-based superconductors *Results in Physics* **18** 103339 (2020)

[68]  C. B. Satterthwaite and I. L. Toepke, Superconductivity of hydrides and deuterides of thorium *Phys. Rev. Lett.* **25**, 741-743 (1970)

**SUPPLEMENTARY INFORMATION**

**Universal Fermi velocity in highly compressed hydride superconductors**

Evgeny F. Talantsev[1,2*]

[1]M.N. Miheev Institute of Metal Physics, Ural Branch, Russian Academy of Sciences,
18, S. Kovalevskoy St., Ekaterinburg, 620108, Russia

[2]NANOTECH Centre, Ural Federal University, 19 Mira St., Ekaterinburg, 620002,
Russia

*corresponding author's E-mail: evgney.talantsev@imp.uran.ru

## S1. Annealed highly-compressed sulphur hydride

Raw $R(T, B_{appl})$ datasets for annealed highly-compressed sulphur hydride ($P = 155$, 160 GPa) reported by Mozaffari *et al* [31] were processed and fitted to Eq. 11. Deduced $B_{c2}(0)$ and $T_c$ values are included in Table I.

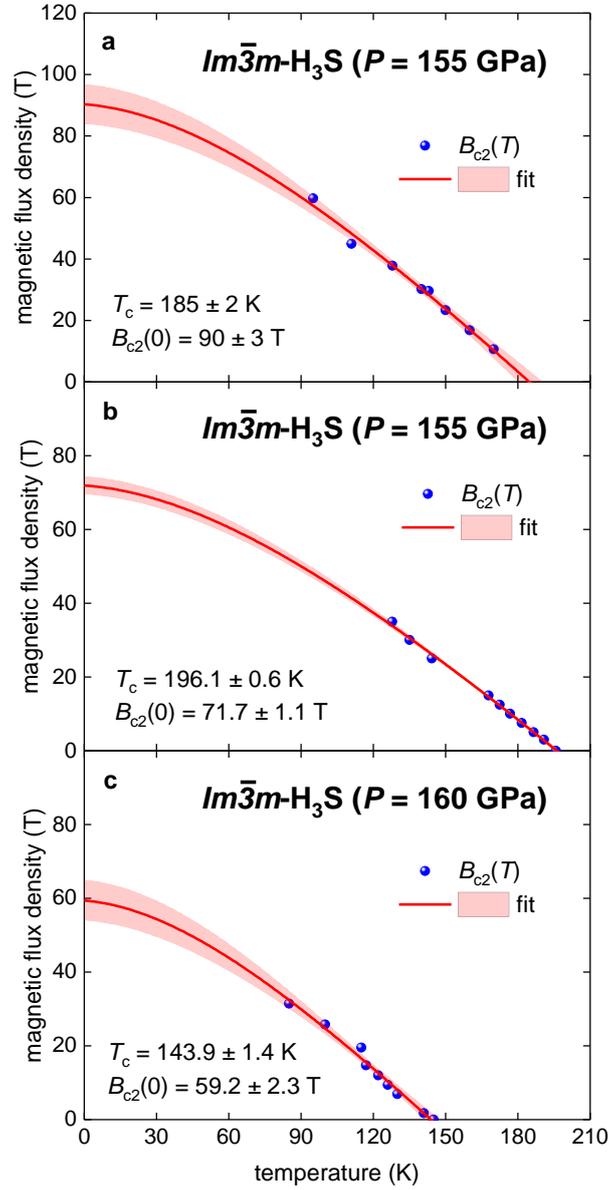

**Figure S1.** The upper critical field, $B_{c2}(T)$, data for highly-compressed $Im\overline{3}m$-H$_3$S phase at $P = 155$ GPa (a,b) and $P = 160$ GPa (c) and data fit to B-WHH model [53] (Eq. 11). Raw $R(T, B_{appl})$ datasets reported by Mozaffari *et al* [31]. Fits quality are (a) $R = 0.9892$; (b) $R = 0.9974$; (c) $R = 0.9837$. 95% confidence bands are shown by pink shaded areas.

## S2. Annealed highly-compressed cerium hydride

Chen *et al* [12] reported on the observation high-temperature superconductivity in superhydrides of cerium. By using the criterion of Eq. 12, we extracted $B_{c2}(T)$ datasets from four $R(T,B_{appl})$ curves reported by Chen *et al* [12] and fitted these $B_{c2}(T)$ datasets to Eq. 11 in Figure S2.

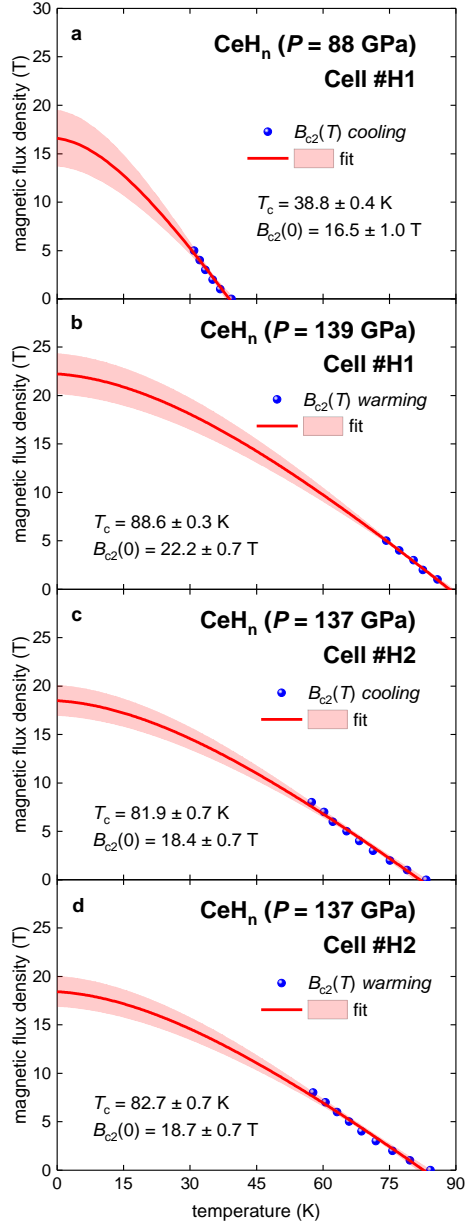

**Figure S2.** The upper critical field, $B_{c2}(T)$, data for highly-compressed superhydrides of cerium and data fits to Eq. 11. Raw $R(T,B_{appl})$ datasets reported by Chen *et al* [12]. (a) Raw $R(T,B_{appl})$ data reported in Fig. S7(a) [12], fit quality is $R = 0.9792$. (b) Raw $R(T,B_{appl})$ data reported in Fig. 1(c) [12], fit quality is $R = 0.9966$. (c,d) Raw $R(T,B_{appl})$ data reported in Fig. 1(d) [12], fit quality is $R = 0.9860$ (c) and $R = 0.9859$ (d). 95% confidence bands are shown by pink shaded areas.

## S3. Annealed highly-compressed cerium hydride

Sun *et al* [54] reported results of magnetoresistance studies for two phases of highly-compressed LaH$_{10}$. By using the criterion of Eq. 12, we extracted $B_{c2}(T)$ datasets for these two phases and fitted these $B_{c2}(T)$ datasets to B-WHH model (Eq. 11) in Figure 4. Deduced $B_{c2}(0)$ and $T_c$ values are included in Table I.

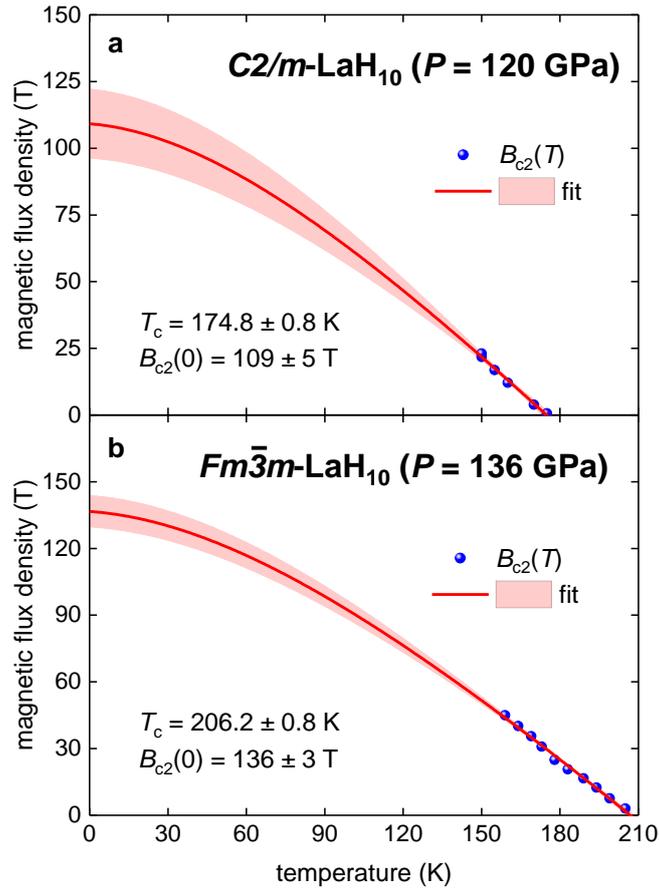

**Figure S3.** The upper critical field, $B_{c2}(T)$, data for highly-compressed LaH$_{10}$ and data fits to Eq. 11. Raw $R(T,B_{appl})$ datasets reported by Sun *et al* [54]. (a) Raw $R(T,B_{appl})$ data reported in Fig. 3(a) [54], fit quality is $R = 0.9907$. (b) Raw $R(T,B_{appl})$ data reported in Fig. 3(b) [54], fit quality is $R = 0.9941$. 95% confidence bands are shown by pink shaded areas.

## S4. Annealed highly-compressed YH₆/YD₆

Recently, Troyan *et al* [4] and Kong *et al* [5] reported on the discovery of new highly-compressed NRTS polyhydrides/polydeuterides of yttrium, $YH_n/YD_n$ (n = 4,6,7,9). Here in Figure 5 we showed extracted $B_{c2}(T)$ for $YD_6$ ($P = 172$ GPa, raw $R(T,B_{appl})$ dataset is from Figure S13(a) [4]) and for $YH_6$ ($P = 200$ GPa, raw $R(T,B_{appl})$ dataset is from Figure S16(a) [4]) and data fits to B-WHH model (Eq. 11). Deduced $B_{c2}(0)$ and $T_c$ are included in Table I.

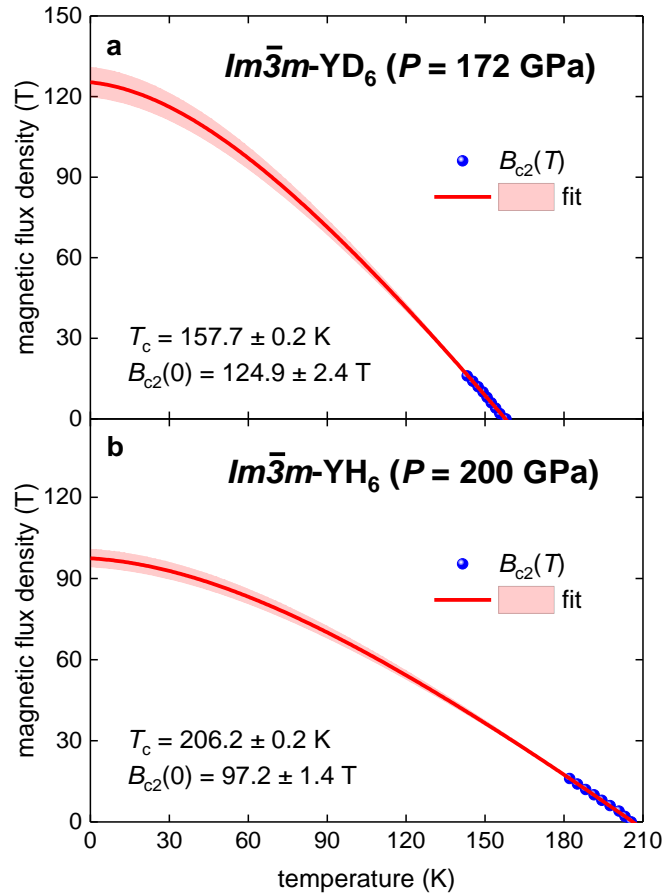

**Figure S4.** The upper critical field, $B_{c2}(T)$, data for highly-compressed YH₆/YD₆ and fits to Eq. 11. Raw $R(T,B_{appl})$ datasets reported by Troyan *et al* [4]. (a) Raw $R(T,B_{appl})$ data reported in Fig. S13(a) [4], fit quality is $R = 0.9971$. (b) Raw $R(T,B_{appl})$ data reported in Fig. S16(a) [4], fit quality is $R = 0.9982$. 95% confidence bands are shown by pink shaded areas.

## S5. Annealed highly-compressed ternary (La,Y)H$_{10}$

Semenok *et al* [8] reported on the discovery of new ternary NRTS polyhydride of (Y,La)H$_{10}$. In Figure 6 we showed extracted $B_{c2}(T)$ datasets for (Y,La)H$_{10}$ phase and data fits to B-WHH model (Eq. 11). Deduced $B_{c2}(0)$ and $T_c$ for this phase are included in Table I.

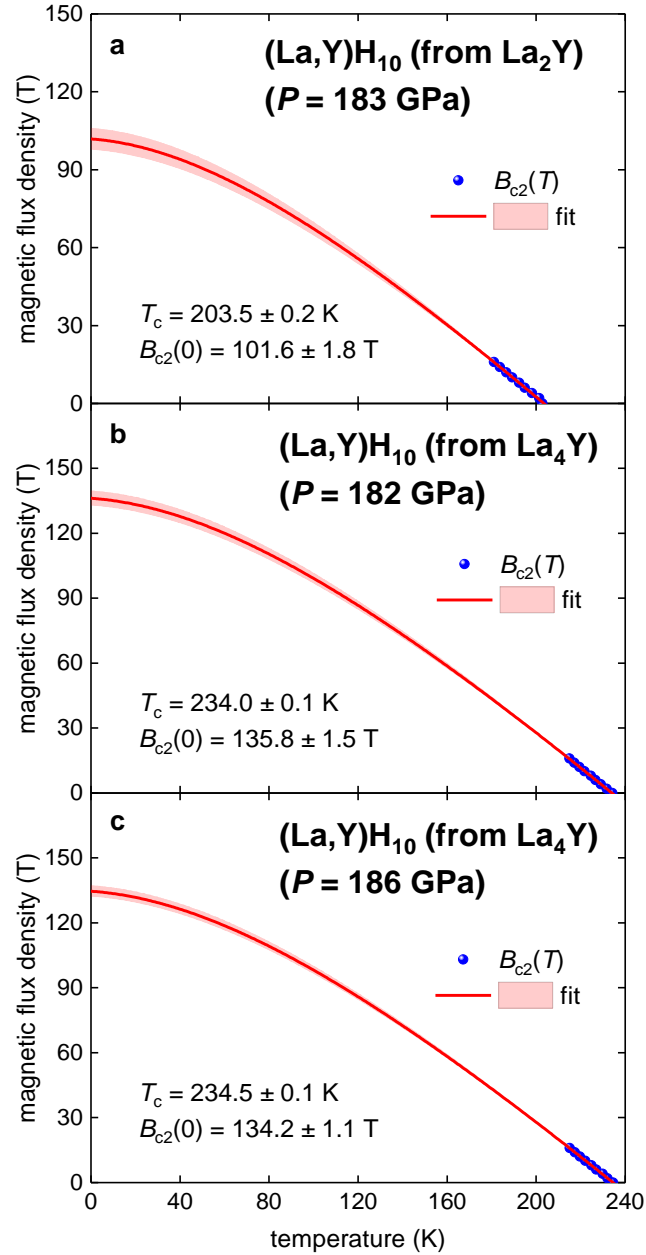

**Figure S5.** The upper critical field, $B_{c2}(T)$, data for highly-compressed (La,Y)H$_{10}$ and fits to Eq. 11. Raw $R(T,B_{appl})$ datasets reported by Semenok *et al* [8]. (a) Raw $R(T,B_{appl})$ data reported in Fig. S27(b) [8], fit quality is $R = 0.9975$. (b) Raw $R(T,B_{appl})$ data reported in Fig. S28(a) [8], fit quality is $R = 0.9991$. (c) Raw $R(T,B_{appl})$ data reported in Fig. S28(a) [8], fit quality is $R = 0.9995$. 95% confidence bands are shown by pink shaded areas.

### S6. Annealed highly-compressed SnH$_{12}$

Recently, Hong *et al* [11] reported on the discovery of a new superconducting polyhydride of *C2/m*-SnH$_{12}$ (*P* = 190 GPa). Extracted $B_{c2}(T)$ datasets for this phase we already reported in our previous work (Table I in Ref. 41). Here in Figure 7 we fitted these datasets to B-WHH model (Eq. 11). Deduced $B_{c2}(0)$ and $T_c$ for this phase are included in Table I.

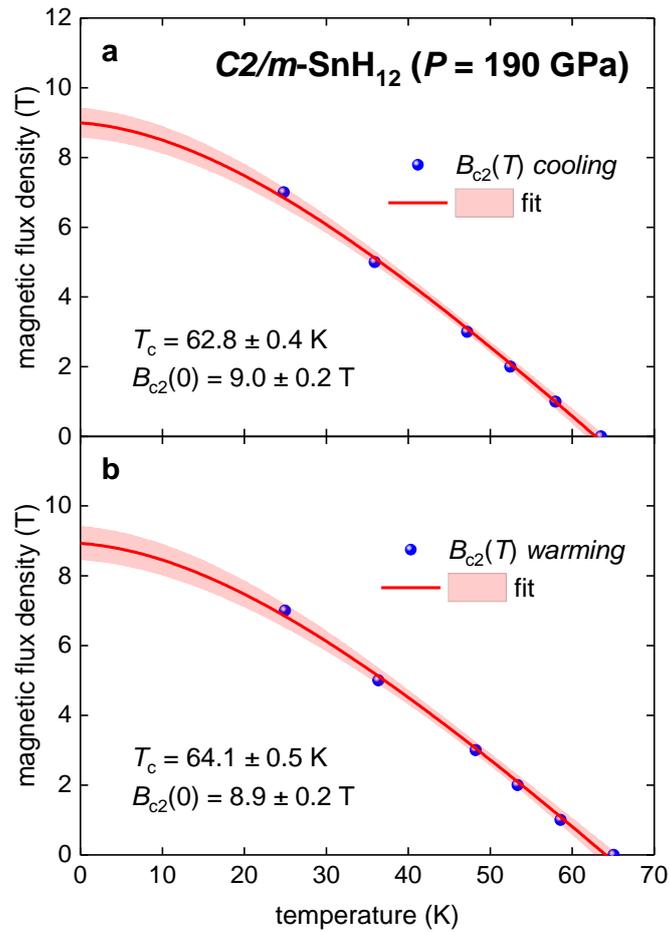

**Figure S6.** The upper critical field data, $B_{c2}(T)$, and data fit to Eq. 11 for *C2/m*-SnH$_{12}$ (*P* = 190 GPa). Raw $R(T,B_{appl})$ datasets reported by Hong *et al* [11] and extracted $B_{c2}(T)$ datasets can be found in Table I in Ref. 41. (a) fit quality is $R = 0.9971$; (b) fit quality is $R = 0.9978$.

## S7. Annealed highly-compressed ThH$_9$ and ThH$_{10}$

Semenok *et al* [16] reported on the discovery of a new NRTS polyhydrides of thorium, ThH$_9$ and ThH$_{10}$. Raw $R(T,B_{appl})$ dataset for mixture of ThH$_9$ and ThH$_{10}$ phases is in Figure 5(c) in Ref. [16]. To deduce $B_{c2}(T)$ dataset for phase ThH$_{10}$ we used the criterion of $R(T, B_{appl})_{criterion} = 4.9\ m\Omega$, while to deduce $B_{c2}(T)$ dataset for phase ThH$_9$ we used the criterion of $R(T, B_{appl})_{criterion} = 0.46\ m\Omega$. In Figure 7 we fitted these datasets to B-WHH model (Eq. 11). Deduced $B_{c2}(0)$ and $T_c$ for this phase are included in Table I.

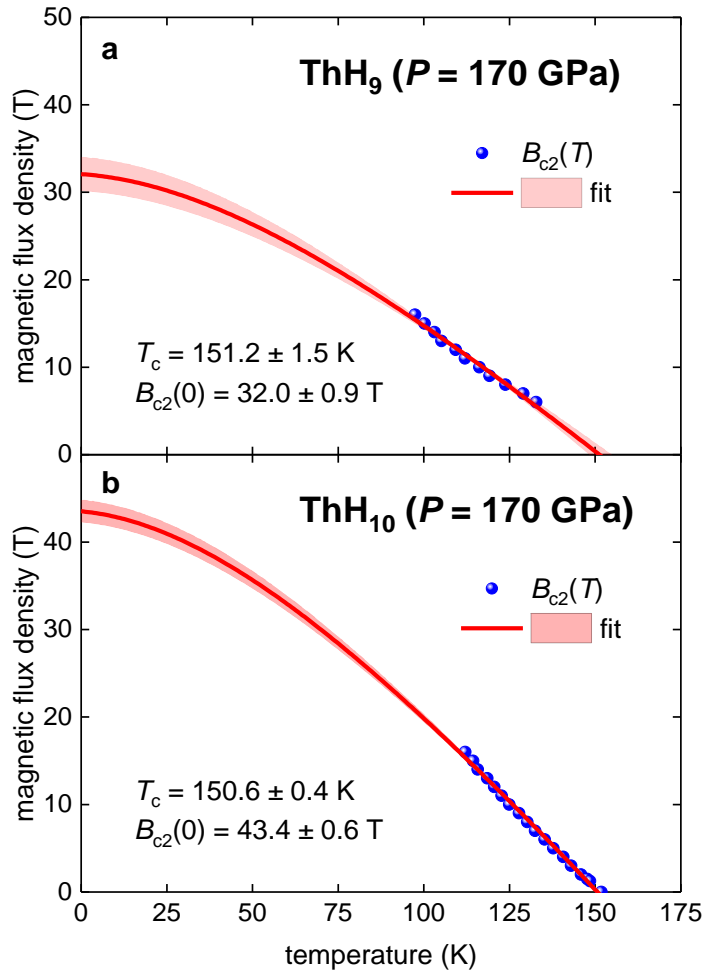

**Figure S7.** The upper critical field data, $B_{c2}(T)$, and data fit to Eq. 11 for NRTS phases of (a) ThH$_9$ and (b) ThH$_{10}$. Raw $R(T,B_{appl})$ datasets reported by Semenok *et al* [16] in their Figure 5(c). (a) fit quality is $R = 0.9866$; (b) fit quality is $R = 0.9957$.